# Astrophysics with heavy-ion beams

P. Senger[1,2] for the CBM Collaboration*

[1] Facility for Antiproton and Ion Research, Darmstadt, Germany

[2] National Research Nuclear University MEPhI, Moscow, Russia

* https://www-cbm.gsi.de/cbmcdb/reportgen.cgi?report=clist-html

E-mail: p.senger@gsi.de

**Abstract**

The "Facility for Antiproton and Ion Research" (FAIR), an international accelerator centre, is under construction in Darmstadt, Germany. FAIR will provide high-intensity primary beams of protons and heavy-ions, and intense secondary beams of antiprotons and of rare short-lived isotopes. These beams, together with a variety of modern experimental setups, will allow to perform a unique research program on nuclear astrophysics, including the exploration of the nucleosynthesis in the universe, and the exploration of QCD matter at high baryon densities, in order to shed light on the properties of neutron stars, and the dynamics of neutron star mergers. The Compressed Baryonic Matter (CBM) experiment at FAIR will investigate collisions between heavy nuclei, and measure various diagnostic probes, which are sensitive to the high-density equation-of-state (EOS), and to the microscopic degrees-of-freedom of high-density matter. The CBM physics program will be discussed.

## 1. Future laboratory experiments on nuclear astrophysics

A major field of research at the future Facility for Antiproton and Ion Research (FAIR) [1] will be nuclear astrophysics with heavy ion beams. In particular, the experiments will cover the production and investigation of extremely rare isotopes, in order to explore the various paths of nucleosynthesis in the universe. Moreover, matter at neutron star core densities will be created and studied in collisions between heavy nuclei, in order to unravel the high-density nuclear matter equation of state, and the microscopic degrees of freedom at these extreme densities. The heavy-ion beams required for these experiments will be provided by the synchrotron SIS100, which accelerates the nuclei up to kinetic beam energies of 11A GeV, while protons reach energies of up to 29 GeV. The unique feature of SIS100 is the production of high-intensity and high-energy heavy-ion beams. An intensity of up to 1010 ions/s is reached, when the ions are fully stripped after pre-acceleration with the existing SIS18, and then injected into SIS100. In this case, beams with energies of up to 11 A GeV will be provided, which will be used for the experiments on high-density QCD matter.

Much higher beam intensities of up to $10^{12}$ ion/s will be achieved, when injecting ions with a lower charge state, for example $^{238}U^{28+}$, into SIS100. In this case the intra-beam scattering, i.e. the Coulomb repulsion between the ions of the beam, is strongly reduced, which permits to accelerate ion beams with higher intensity. This operation mode requires an extremely good vacuum better than $5 \cdot 10^{-12}$ mbar inside the beam pipe, in order to avoid the ionization of the beam particles, and, hence, the loss of the beam. Such conditions can be achieved by cooling the beam pipe down to temperatures of 12 K, and using most of the beam pipe as a cryogenic pump. Because of the lower charge state of the beam, only beam energies of up to 2 A GeV can re reached. However, this is sufficient to provide high-intensity secondary beams of rare isotopes, which will be produced and separated in-flight by a large-acceptance Superconducting Fragment Separator (SFRS), and investigated downstream the SFRS by various experimental setups of the NUSTAR collaboration (**Nu**clear **St**ructure, **A**strophysics and **R**eactions).

The beam energies delivered by the SIS100 synchrotron at FAIR are also well suited to create dense nuclear matter in the reaction volume of heavy-ion collisions. Event generators, which are used for microscopic simulations of such reactions, predict, that the nucleons of the two colliding nuclei pile up to 5 times the normal nuclear matter density in central Au+Au collision at already a kinetic beam energy of 5A GeV, and to even higher densities at 11A GeV [2]. Above 5 times nuclear matter density, i.e. under conditions which may be realized in the core of neutron stars, the nucleons will start to overlap, and the elementary building blocks of matter, quarks and gluons, may roam freely in the reaction volume [3]. The exploration of QCD matter under such conditions is a central part of the research program of the Compressed Baryonic Matter (CBM) experiment. FAIR comprises two more scientific pillars, the PANDA and the APPA experiments. The PANDA experiment (Anti**P**roton **An**nihilation at **Da**rmstadt) is located at the High-Energy Storage Ring (HESR), and is designed to perform a rich research program in hadron physics using high-energetic antiproton beams. Also experiments with low-energy antiproton beams are under consideration, in order to produce and investigate anti-hydrogen atoms. Such experiments may shed light on the symmetry breaking between matter and antimatter in the early universe right after the big bang. APPA is an umbrella for several collaborations working on **A**tomic **P**hysics, **P**lasma physics and **A**pplied research in biophysics, medical science, and material sciences. The APPA collaborations share installations and experimental techniques. The FAIR project, including civil construction, accelerator construction, and the realization of the four experimental pillars is progressing, and first experiments are planned for 2025.

## 2. Unraveling the origin of elements in the universe

The different paths of nucleosynthesis along the chart of isotopes are illustrated in figure 1 together with the relevant astrophysical production sites. Nuclei up to the mass of iron are created by fusion processes which take place in high-mass stars. These nuclei are ejected into space by supernova explosions, providing the seed for the next generation of stars. Binary systems of suns and neutron stars provide high proton fluxes, which are needed for the synthesis of isotopes with proton excess via the rapid proton capture (rp) and proton capture (p). In stars with very large masses, about 50% of the heavy elements up to Bismuth are produced by slow neutron capture (s-process) by stable nuclei. The other 50% of heavy nuclei including the trans-actinides are formed by rapid neutron captures followed by beta decays. This r-process requires extremely high neutron fluxes, as expected to occur in core-collapse supernovae or in neutron star mergers. Because of the very short life-times and their extremely low production probability in heavy-ion collisions, the isotopes potentially participating in the r-process have been rarely studied in the laboratory. New information on the possible site of the r-process was obtained from the multi-messenger signals, which followed the measurement of gravitational waves generated by a neutron star merger [4,5]. The observed electromagnetic spectrum agrees very well with the predicted "kilonova" signal, which is generated by radioactive decays of heavy elements produced by the intense neutron flux of the merger [6]. Therefore, most of the heavy elements including gold, platinum and uranium might be synthesized in star mergers.

As illustrated in figure 1, the rp, p and r-processes involve isotopes with a large deficit or excess of neutrons, which are short-lived. In order to produce and investigate such nuclei in the laboratory, both high-intensity nuclear beams and extremely selective experimental devices are required. In order to retrace, for example, the path of the r-process along the chart of nuclei, information on the nuclear properties is needed, such as the neutron separation energies (masses) and beta half-lives. The Super-FRS at FAIR is designed as large-acceptance in-flight separator for the precise selection of rare isotopes, which will be produced and spatially separated within some hundred nanoseconds. This method allows to identify also very short-lived nuclei, which will be first selected and then guided to subsequent detector setups or storage rings, where both masses and half-lives of rare isotopes of all elements up to uranium will be determined.

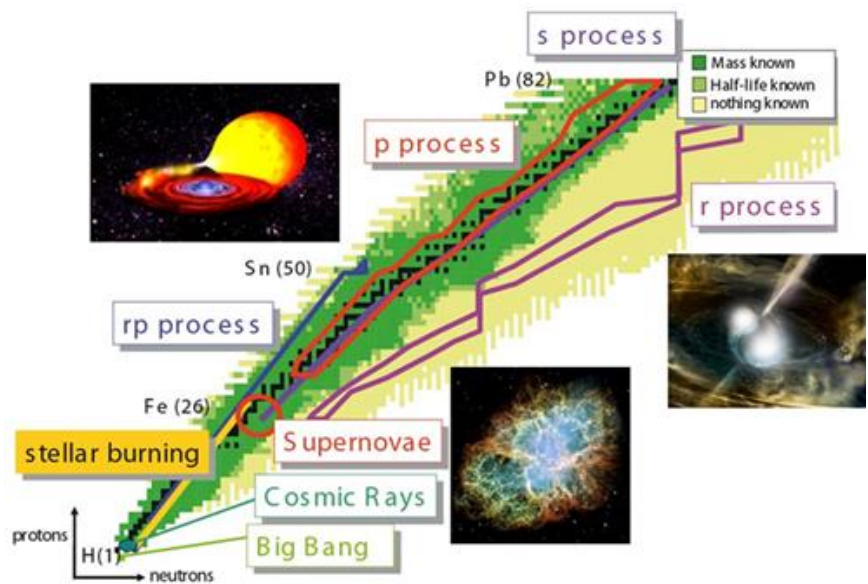

Fig.1: Illustration of the nucleosynthesis processes in the universe along the chart of nuclei. Elements up to iron are created by nuclear fusion processes in high mass stars. About half of the heavy elements up to Bismuth are created by slow neutron capture (s-process) in low and midsize stars. The other half of the heavier elements including the trans-actinides are produced by rapid neutron capture (r-process) in the high neutron fluxes generated in supernova explosions and neutron star mergers. The rapid proton capture (rp-process) and the proton capture (p-process) takes place in binary systems of a sun and a neutron star.

Figure 2 depicts the configuration of the superconducting magnets of the SFRS, together with trajectories of particles with different mass/charge ratio. In the left lower panel the isotopes are shown, which are produced with almost the same velocity in the reaction of a 1.1 A GeV $^{238}$U beam with a 4 g/cm$^2$ C target. They hit the first Al degrader, where the nuclei suffer an energy loss depending on their position. The center lower panel depicts the isotopes, which have been selected after the pre-separator. In the right lower panel the isotopes are shown, which have been selected by the main separator after a second energy loss, demonstrating the selection power for the $^{132}$Sn isotope.

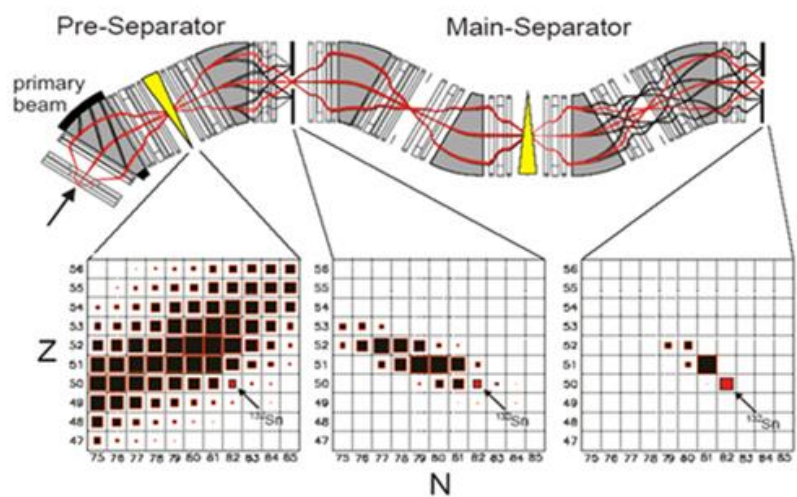

Fig. 2: Upper panel: Sketch of the magnet configuration of the Super-Fragment Separator with two Al degraders (yellow wedges), together with particle trajectories with different mass/charge ratio. The lower left panel depicts the isotopes, which are produced by a 1.1 A GeV $^{238}$U beam hitting a 4 g/cm$^2$ C target, and reach the first Al degrader, where they suffer an energy loss depending on their position. The centre lower panel depicts the isotopes which passed the pre-separator. After the main separator including a second energy loss, the $^{132}$Sn isotope is clearly identified.

Downstream the Super-FRS, the rare isotope beams can be distributed into three branches, which serve different experimental areas including a new storage-ring complex. Spectroscopy experiments will be performed after slowing down the secondary beams, and high-energy rare isotope beams are used for reaction experiments.

In conclusion, the Super-FRS and its subsequent detection facilities are designed to generate secondary beams of rare, short-lived isotopes with high purity, and to measure their masses and lifetimes with high precision. These experiments will scout a yet unknown territory of the nuclear chart, and substantially contribute to the exploration of the various paths of nucleosynthesis in the universe.

### 3. Nuclear matter at neutron star core densities

The recent observation of gravitational waves emitted from a neutron star merger triggered extensive theoretical efforts, in order to simulate the signal generation and the underlying collision dynamics. An example is shown in figure 3, which depicts the distribution of the rest mass density (upper left panel), and the temperature profile in the equatorial plane at a post merger time of 6.34 ms (upper right panel) [7]. The density reaches values of up to 4 times saturation density $\rho_0$, whereas in some hot spots the matter is heated up to about 50 MeV. For comparison, a semi-central collision of two gold nuclei has been calculated with the UrQMD event generator for a beam kinetic energy of 1.5A GeV. The results are shown in the lower panel of figure 3. In this case, densities of up to 3 $\rho_0$ (left panel) and temperatures of up to 100 MeV (right panel) are reached. It is worthwhile to note, that the nuclear matter densities and the temperatures predicted for neutron star mergers and for heavy-ion collisions are quite comparable.

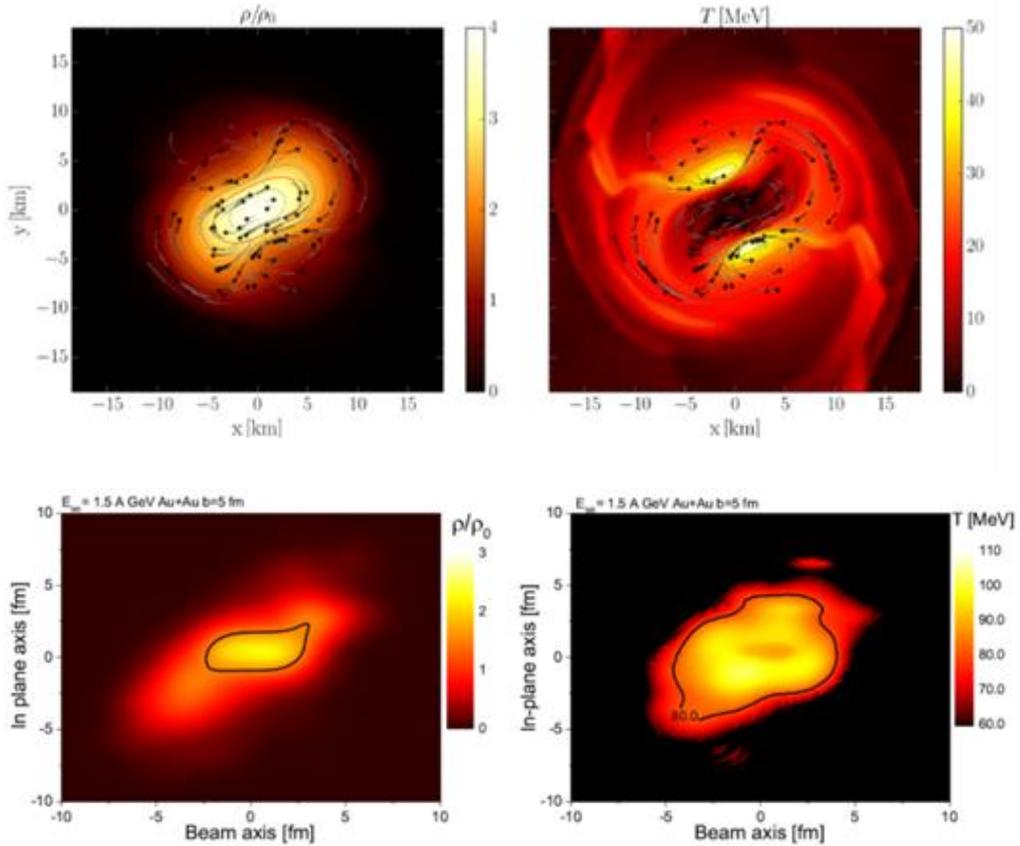

Fig.3: Distribution of the density $\rho$ in units of $\rho_0$ (left panel) and of the temperature (right panel) in the equatorial plane of a neutron star merger event, plotted at a post-merger time of t= 6.34 ms. Lower panels: Distribution of the net-baryon density (left) and of the temperature (right) in the reaction plane of a semi-central (b= 5 fm) Au+Au collision a beam energy of $E_{lab}$= 1.5A GeV as calculated with the UrQMD transport model. Taken from [7].

Despite the substantial differences in size, lifetime, and isospin, the neutron star merger and the fireball created in a heavy-ion collision as shown in figure 3 have been simulated using the same equation-of-state (EOS) of nuclear matter. This opens the unique possibility, to study this fundamental property of elementary strongly interacting matter both by astrophysical observations, and in the laboratory via heavy-ion collisions. These complementary methods should lead to a conclusive idea of the properties of dense QCD matter. For example, information on the high-density EOS can be deduced from a frequency analysis of the gravitational waves emitted by a neutron star merger, or by the simultaneous measurement of mass and radius of neutron stars, as planned by the NICER experiment [8]. In the laboratory, the EOS can be extracted from the collective flow of particles generated in a collision between heavy nuclei, or from the yield of strange particles measured at in heavy-ion collisions at subthreshold bombarding energies.

Baryon densities of 5 $\rho_0$ and higher are expected to exists in the core of heavy neutron stars [9]. At these densities, the nucleons are expected to melt, and the fundamental degrees of freedom emerge, presumably forming a mixed phase of hadrons and quarks [3,10]. Also in central collisions between two gold nuclei at a beam kinetic energy of 5A GeV, net baryon densities of up to 5 $\rho_0$ are produced, and even densities of 8 $\rho_0$ are expected to be reached in the centre cell of the fireball created at energies of 10A GeV [2]. Therefore, the QCD phase at neutron star core densities can also be studied in heavy-ion collisions.

In this section, the status and perspectives of laboratory experiments on the determination of the nuclear matter equation of state are reviewed. Moreover, the prospective exploration of the QCD phase diagram at high densities will be outlined, including the search for a possible phase transition between hadronic and partonic matter. Finally, heavy-ion experiments will be proposed, to shed light on the hyperon puzzle in neutron stars.

### 3.1 The high-density equation of state

The equation of state (EOS) of nuclear matter relates pressure, density, volume, temperature, energy, and isospin. For cold matter, the EOS can be expressed as energy per nucleon, which then depends on baryon density and isospin asymmetry. Various EOS are shown in figure 4. The lower curves describe the EOS for symmetric nuclear matter. In this case, the EOS exhibits a minimum at saturation density $\rho_0$, with a binding energy of -16 MeV, and with a curvature, defined as nuclear incompressibility $K_{nm}$. The upper curves correspond to EOS for pure neutron matter.

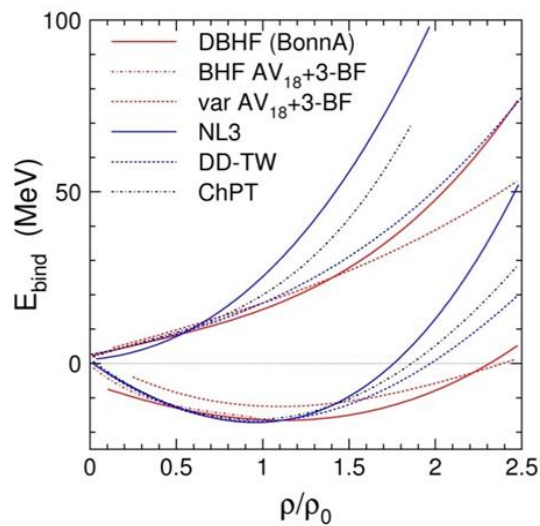

Fig.4: EOS for isospin-symmetric nuclear matter (lower curves) and for neutron matter (upper curves). [12]. The difference of the two set of curves at saturation density $\rho_0$ corresponds to the symmetry energy.

The nuclear incompressibility at saturation density has been extracted from experiments on giant monopole resonances of heavy nuclei, and was determined to $K_{nm}(\rho_0) = 230\pm10$ MeV [11]. Heavy-ion experiments at GSI have extended our knowledge on the EOS of symmetric nuclear matter up to densities of about 2 $\rho_0$. An important observable sensitive to the EOS is the collective flow of particles, which driven by the pressure gradient in the compressed reaction volume.

A systematic study of the collective particle flow has been performed by the FOPI collaboration for Au+Au collisions at beam kinetic energies from 0.4A to 1.5A GeV. FOPI measured the flow of protons, deuterons, tritons and 3He perpendicular to the reaction plane [13]. This particular elliptic flow pattern, the so called squeeze-out, is formed by the spectator fragments of the nuclei, which shadow the flow component into the reaction plane, as illustrated in figure 5.

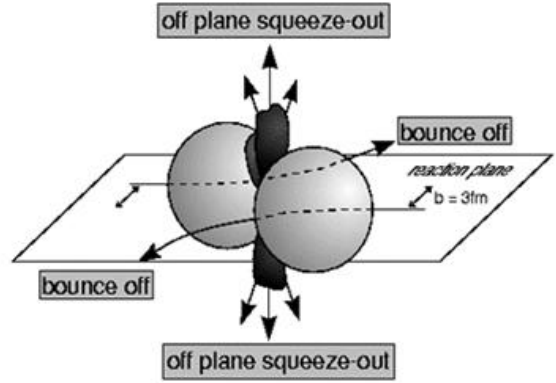

Fig.5: Illustration of the in-plane flow ("bounce off") and of the off-plane flow ("squeeze-out") for a non-central heavy-ion collision at intermediate beam energies.

The experimental data have been compared to results of IQMD transport calculations, which could explain the data best when assuming a soft EOS ($K_{nm}$=190 ± 30 MeV) and taking into account momentum-dependent interactions. The comparison of experiment and simulation is shown in figure 7, which depicts the results of IQMD calculations with momentum dependent interactions for a hard EOS ($K_{nm}$=380 MeV, red curve) and for a soft EOS ($K_{nm}$=200 MeV, blue curve), together with the experimental data (yellow area).

Another observable sensitive to the EOS for symmetric nuclear matter is particle production at subthreshold beam energies. This method has been pioneered by the KaoS collaboration at GSI, which studied subthreshold K+ production in heavy ion collisions. In nucleon-nucleon collisions, the minimum kinetic beam energy required to create a $K^+$ meson in a p+p $\rightarrow$ $K^+\Lambda$p reaction is 1.6 GeV. In collisions between heavy ions, however, $K^+$ mesons can also produced in subsequent collisions involving nucleons, $\Delta$ resonances, and pions, where the energy of multiple collision is accumulated. These multiple collisions happen more frequently, if the baryon density in the reaction volume is high. Consequently, the $K^+$ meson yield increases with increasing baryon density, and if high densities are reached, the EOS should be soft. However, also effects like the Fermi energy of the nucleons, in-medium modifications of mesons, and short-range correlations may contribute to subthreshold particle production [14]. Therefore, it is necessary to investigate in addition to collisions between heavy nuclei also collisions between light nuclei, which serve as a reference system. These light systems are much less compressed than heavy systems, and, hence, are much less sensitive to the EOS, but incorporate the in-medium effects mentioned above.

Figure 6 depicts the excitation function of $K^+$ production measured in Au+Au and C+C collisions at subthreshold beam energies [15], together with results of various transport models for a hard and a soft EOS[16,17]. In figure 6, the data and the simulation results are plotted as double ratios, i.e. the K+ yield per nucleon in Au+Au collisions is divided by the $K^+$ yield per nucleon in C+C collisions. This ratio does not only correct for the above mentioned in-medium effects, but also for systematic errors of the data and the simulations. The strong increase of the data towards lower beam energies can be reproduced by the RQMD and IQMD model calculations with momentum dependent interactions and in-medium potentials, and assuming a soft EOS ($K_{nm}$=200 MeV), whereas the simulation results with a hard EOS ($K_{nm}$= 380 MeV) show only a weak beam energy dependence.

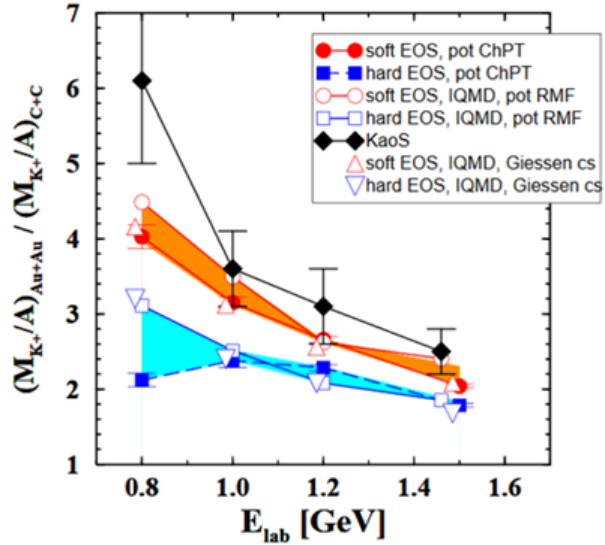

Fig. 6: Ratio of the K$^+$ meson multiplicity per participant nucleons from Au+Au collisions over the one from C+C collisions as function of beam energy. Black symbols: KaoS data [15]. The data are compared to transport simulations with momentum dependent interactions for a soft EOS ($K_{nm}$=200 MeV, red symbols) and for a hard EOS ($K_{nm}$=380 MeV, blue symbols). Taken from [12].

The results of the experiments at GSI on the EOS of symmetric nuclear matter up to baryon densities of about twice saturation density are summarized in figure 7. The results of the flow and kaon measurements by the FOPI and KaoS experiments, respectively, are represented by the yellow area, while the results of IQMD transport calculations for the hard EOS ($K_{nm}$= 380 MeV) are described by the red line, and for the soft EOS ($K_{nm}$= 200 MeV) as blue line.

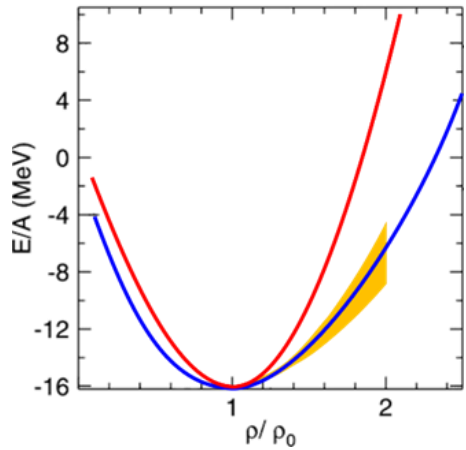

Fig. 7: EOS for symmetric nuclear matter as function of baryon density. The lines represent results of IQMD calculations with momentum dependent interactions [13] for a hard EOS ($K_{nm}$=380 MeV, red line) and for a soft EOS ($K_{nm}$=200 MeV, blue line). Yellow area: Results of FOPI and KaoS experiments from Au+Au collisions [13, 15].

Up to now the EOS of symmetric nuclear matter up to twice saturation density has been discussed. However, in order to contribute to our understanding to neutron stars with twice the solar mass, the EOS should be studied at substantially higher densities. Various EOS for neutron stars are presented in figure 8, which has been used to calculate the mass and the central density of the star via the Tolman–Oppenheimer–Volkoff (TOV) equation [18]. In order to explain the most massive known neutron stars, such as the millisecond pulsar PSR J0740+6620 with a mass of 2.14+0.10−0.09 solar masses [19], or PSR J1614-2220 with a mass of 1.97±0.04 solar masses [20], densities above about 4 $\rho_0$ have to be assumed.

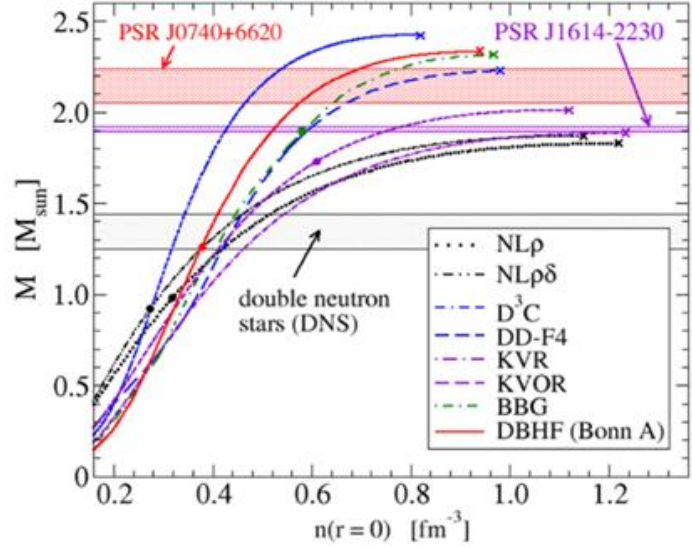

Fig. 8: Neutron star mass as function of the central density for different EOS as calculated by the Tolman–Oppenheimer–Volkoff (TOV) equation [18].

Heavy-ion experiments at the AGS in Brookhaven have pioneered the investigation of the equation-of-state of dense symmetric nuclear matter by measuring both the transverse (in-plane) flow and the elliptic flow of protons in Au+Au collisions at energies between 2A GeV and to 11A GeV [20]. Compared to relativistic transport model calculations, the transverse flow data can be fairly well reproduced by a soft EOS ($K_{nm}$ = 210 MeV), while the data on elliptic flow suggest a stiff EOS ($K_{nm}$ = 300 MeV) [21]. The resulting constraint from this analysis for the high-density EOS for symmetric matter, expressed as pressure versus baryon density, is depicted in figure 9 as the grey-hatched area. For comparison, also the simulation results are shown for the hard and the soft EOS mentioned above. The yellow area corresponds to the FOPI [13] and KaoS [15] results.

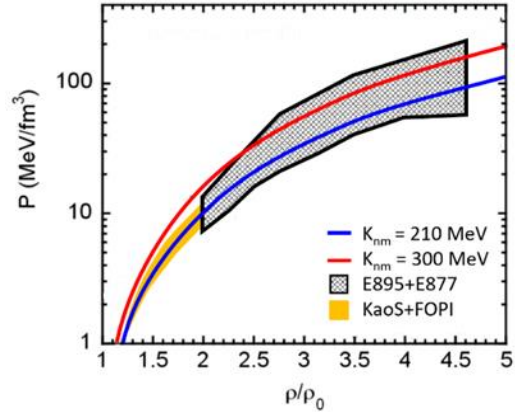

Fig. 9: The EOS for symmetric nuclear matter, expressed as pressure versus baryon density. Grey hatched area: constraint extracted from proton flow data taken at the AGS [21,22]. Yellow area: Constraint from fragment flow and kaon date taken at GSI [13,15]. Red line: hard EOS, blue line: Soft EOS [21].

In conclusion, the flow and kaon data measured in heavy-ion collisions at beam energies up to 1.5A GeV consistently can be explained by transport models assuming a soft EOS for symmetric matter up to baryon densities of 2 $\rho_0$. At higher densities, i.e. for beam energies above 2A GeV, the interpretation of the existing flow data by transport models is compatible with both hard and soft equations of state, only ruling out extremely soft or extremely high values for the nuclear incompressibility.

It should be noted, that the elliptic flow pattern depends on the beam energy. At beam kinetic energies from about 100 MeV to 4A GeV, the elliptic flow is generated by particles which are "squeezed-out" perpendicular to the reaction plane (see figure 5). Around 4A GeV the elliptic flow vanish, and changes is direction into the reaction plane, because the flow particles are no longer shadowed by the spectators. Moreover, the compressed, almond shaped fireball expands stronger into the direction of the reaction plane, driven by the pressure gradient, which converts the initial spatial anisotropy into the observed momentum distribution of identified particles. Up to now, only scarce experimental information on the elliptic flow is available at beam energies, where high net baryon densities are created, i.e. between 2A and 10A GeV. Moreover, the expected flow values are rather small. Therefore, future precision flow measurements are required, to improve the data situation, and our knowledge on the high-density EOS. Valuable information will be provided by the determination of the collective flow of a variety of identified particles, including light fragments and hyperons. Such measurements will be performed by future experiments, like the Compressed Baryonic Matter experiment at FAIR.

Complementary to the flow measurements, subthreshold particle production may help to probe the high-density EOS, similar to the kaon measurements at lower beam energies and lower baryon densities. In the beam energy range of interest, i.e. up to about 10A GeV, multi-strange (anti-) hyperons would be the favorable candidates, as their production thresholds correspond to beam energies ranging from 3.7 GeV to 12.7 GeV (see table 1).

Table 1: Production processes of multi-strange (anti-) hyperons, and their threshold kinetic beam energy

| Production process | $E_{thr}$ (GeV) |
|---|---|
| $pp \to \Xi^- K^+ K^+ p$ | 3.7 |
| $pp \to \Omega^- K^+ K^+ K^0 p$ | 7.0 |
| $pp \to \Lambda^0 \text{anti-}\Lambda^0 pp$ | 7.1 |
| $pp \to \Xi^+ \Xi^- pp$ | 9.0 |
| $pp \to \Omega^+ \Omega^- pp$ | 12.7 |

Like the kaons at low beam energies, multi-strange hyperons can be also produced via multistep collisions including strangeness exchange reactions, with kaons, lambda and sigma hyperons involved [23,24]. Figure 10 illustrates the production mechanisms for $\Omega^-$ hyperons in central Au+Au collisions at a beam energy of 4A GeV, as simulated with the HYPQGSM transport model [25]. By far most of the $\Omega^-$ hyperons are created in collisions of two hyperons, preferably by a strangeness exchange reaction involving a lambda or a sigma hyperon and a Xi-hyperon. Less abundant but still notable are reactions between kaons and hyperons. The number of these multistep processes, and, hence, the yield of multi-strange hyperons increases with increasing density in the fireball, which is sensitive to the EOS.

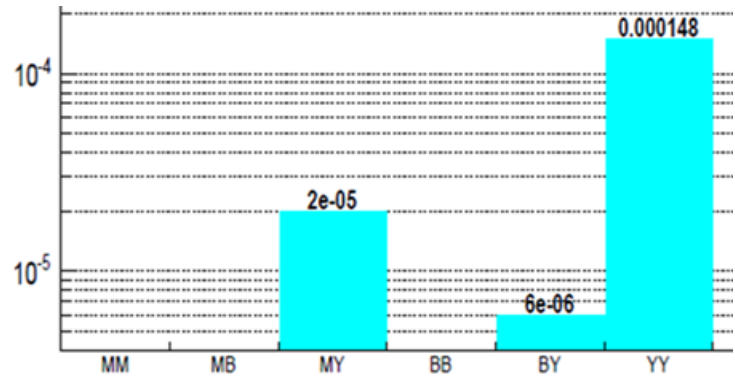

Fig. 10: Multiplicity of $\Omega^-$ hyperons produced in central Au + Au collisions at beam kinetic energy of 4A GeV via different reaction mechanisms according to the HYPQGSM event generator. MY: meson-hyperon, BY: Baryon-hyperon, YY: Hyperon-hyperon [25].

The production of multi-strange (anti-)hyperons in central Au+Au collisions at a beam energy of 4A GeV has been also simulated using the PHQMD transport model with different EOS [26]. Preliminary results of the calculations indicate, that multi-strange (anti-) hyperons are more abundantly produced when assuming a soft EOS as compared to a hard EOS. It was further observed, that this difference in yields increases with the mass of the hyperons. This shows, that the EOS effect on the particle yield is more pronounced, the more the particle is produced subthreshold. In conclusion, the measurement of the excitation function of hyperon production in collisions of heavy ions and in light reference systems, will be a relevant component of the CBM research program, in order to determine the high-density EOS of nuclear matter.

As indicated in figure 4, one has to measure also the symmetry energy $E_{sym}$, in order to determine the EOS relevant for neutron stars. The value of the symmetry energy at saturation density indicated in figure 4 has been determined by experiments with isospin-asymmetric nuclei [27]. In order to determine $E_{sym}$ at higher densities, the ASY-EOS collaboration performed measurements of the elliptic flow of neutrons and protons in Au+Au collisions at a beam energy of 0.4A GeV, where densities up up to 2 $\rho_0$ can be reached [28]. When analysing the experimental results with an UrQMD transport code, a value of $E_{sym}$=55 ± 5 MeV was extracted.

A complementary method to extract the symmetry energy in heavy ion collisions is to study the yields of particles with opposite isospin, for example with $I_3 = ±1$, which reflect the density of neutrons and protons. A promising observable might be the $\Sigma^-$(dds)/$\Sigma^+$(uus) ratio, which can be related to the n(ddu)/p(uud) ratio. The experimental challenge is to identify $\Sigma$ hyperons, as one of the daughters is always a neutral particle. However, with a high-resolution tracking detector, $\Sigma$ hyperons can be reconstructed via the missing mass method. However, the question is, whether the difference between symmetric nuclear matter and neutron matter still matters at the very high densities in the core of neutron stars, as at baryon densities well above 5 $\rho_0$ the matter properties might be dominated by the strong interaction, and β-equilibrium might be of less importance [29].

The EOS of nuclear matter at high baryon densities depends on the role of multi-body interactions between the nucleons, which are poorly known. For example, with increasing density of neutron star matter, the neutron chemical potential also increases rapidly, until it exceeds the chemical potential of hyperons. In such a situation, the neutrons start to decay weakly into hyperons, thus reducing the pressure, and softening the EOS. This effect limits the mass of neutron stars to values well below 2 solar masses, which is in contradiction with observations. On the other hand, the appearance of hyperons at high density seems to be unavoidable. Whether or not this "hyperon puzzle" exists, depends strongly on the strength of the NY, NNY, and NYY interactions, with N the neutron and Y the hyperon. For example, in chiral effective field theory, which takes into account only hadronic degrees-of-freedom, lambda condensation is prevented up to densities of about 5 $\rho_0$ by assuming repulsive 3-body ΛNN interactions. [30]. Calculations based on the Quark–Meson-Coupling model predict hyperons to appear in neutron stars above densities of 3 $\rho_0$, and prohibit the softening of the EOS also by 3-body forces [31].

In order to provide new information on the ΛN, ΛNN and ΛΛN interactions, the CBM experiment will measure life-times and binding energies of hypernuclei, including double-lambda hypernuclei. Calculations with a UrQMD-hydro hybrid model and with a hadronic cascade model predict a maximum of the production probability of light (double-) lambda hypernuclei in heavy ion collisions at FAIR energies [32]. This maximum originates from an interplay of the hyperon production, which rises with increasing beam energy, and the production of light nuclei, which decreases with increasing beam energy.

## 3.2 Phases of QCD matter at high densities

Calculations on the structure of neutron stars [10] and on the dynamics of neutron star mergers [33] indicate, that above baryon densities of 4 - 5 $\rho_0$ the nucleons may start to melt, and the elementary building blocks of matter, quarks and gluons, are set free. Whether this transition from hadronic to partonic matter is of first order or a continuous process is still a matter of debate. In contrast, for very high temperatures and vanishing baryon chemical potential, a smooth crossover from the hadronic phase to the quark-gluon plasma is predicted by Quantum Chromo Dynamics (QCD), the fundamental theory of strong interaction. According to these lattice QCD calculations, this transition is characterized by a pseudo-critical temperature of about 155 MeV [34,35]. Recently, lattice QCD calculation have been extended non-zero baryon chemical potentials, and found that in the chiral limit the critical temperature for a possible chiral phase transition must be lower than Tc0 = 132 + 3 − 6 MeV [36]. However, QCD still fails to make predictions for large baryon chemical potentials, where effective models suggest structures like first order transitions with critical endpoints, or phases like quarkyonic matter [37,38].

Models based on the concept of quark-hadron continuity predict the gradual emergence of quark degrees of freedom with increasing density, together with a partial restoration of chiral symmetry [3,29]. In this case, there would be no first order transition at high densities and low temperatures. An example of such a model is illustrated in figure 11, which sketches three configurations of baryonic matter at different density [29]. The model describes nucleons with a hard core and a pion cloud. At saturation density, the nucleons are well separated. At baryon densities between 2 and 4-5 $\rho_0$, the pion clouds overlap, and the quarks in the clouds start to move freely within the overlap zone. At high densities, when the nucleon cores touch, also the most of the constituent quarks get liberated. In the lower panel, the corresponding quark occupation function is sketched for different momenta, illustrating the fraction of free quarks in blue and the bound quarks in red. This scenario describes a continuous phase transition from nuclear to quarkyonic matter. The model aims at the description of dense QCD matter in massive neutron stars, and provides a soft EOS at densities of 2-4 $\rho_0$, where the pion clouds overlap, and a stiff EOS at higher densities, as required for the stability of neutron stars with twice the solar mass.

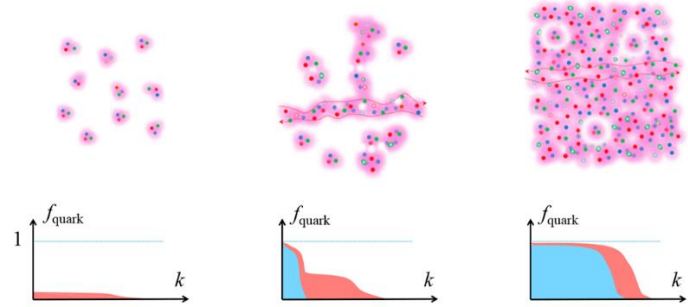

Fig. 11: QCD matter degrees-of-freedom at different densities together with the quark occupation functions versus momentum (see text) [29].

The available experimental information on the QCD phase diagram is presented in figure 12, which depicts the chemical freeze-out temperature of the fireball produced in heavy-ion collisions, as function on the baryon chemical potential [39]. These parameters are obtained by a fit of the statistical hadronization model to the particle yields measured at different collision energies. The dotted line represents a calculation based on a fixed energy per nucleon of 1 GeV [40].

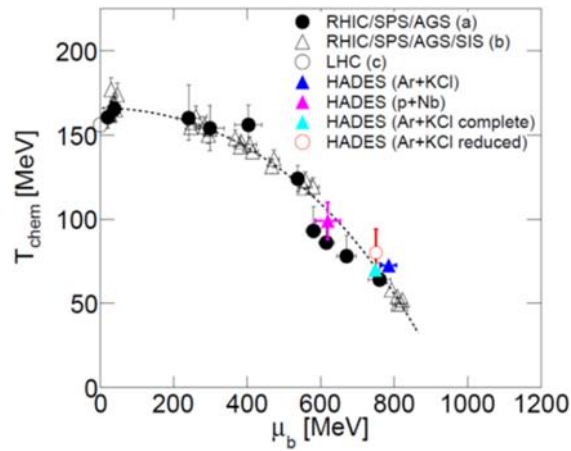

Fig.12: Chemical freeze-out temperatures $T_{chem}$ as function of the baryon chemical potential $\mu_b$ extracted from particle yields measured at different energies [39]. The black dashed line is calculated assuming a fixed energy per nucleon of 1 GeV [40].

At the highest beam energies, provided by the Large Hadron Collider (LHC) at CERN in Switzerland, the ALICE collaboration determined a freeze-out temperature of 156 MeV at $\mu_b = 0$ [41]. This value perfectly agrees with the pseudo-critical temperature calculated by lattice QCD. Considering the QCD prediction for the upper limit of a possible chiral phase transition temperature of $T_c^0 = 132 + 3 - 6$ MeV [36], the corresponding baryon chemical potential should be larger than $\mu_b = 500$ MeV, and the most promising beam energy to search for the critical point would be well below $\sqrt{s_{NN}} = 6\text{-}7$ GeV.

## 3.3 Exploring hot and dense QCD matter with heavy-ion collisions

In this section, the most promising observables are discussed, which are sensitive to the different phases of hot and dense QCD matter as created in the reaction volume of energetic heavy-ion collisions.

3.3.1 The elliptic flow of identified particles

An important observable of the hot and dense matter created in heavy-ion collisions is the elliptic flow of identified particles. The experimental finding, that the elliptic flow increases by 50% from to top SPS energy ($\sqrt{s_{NN}} = 17.2$ GeV) to top RHIC energy ($\sqrt{s_{NN}} = 200$ GeV), was taken as indication for the perfect liquid dynamics of the Quark-Gluon Plasma (QGP). In addition, a universal scaling of the elliptic flow of different hadrons was found: when dividing the elliptic flow signal $v_2$ by the number of participant quarks $n_q$ of each hadron, and plotting it as function of the transverse kinetic energy per quark ($m_T-m_0$)$n_q$, all particles fall on the same curve. This observation was taken as indication, that the elliptic flow was generated in the early QGP phase of the collision. Therefore, the STAR collaboration measured the excitation function of the elliptic flow of identified particles [42], and searched for the energy where the universal scaling vanishes, indicating the disappearance of the QGP phase, and possibly the location of a critical point. It was found that the scaling holds down to the lowest possible energy of $\sqrt{s_{NN}}=7.7$ GeV, where the luminosity of the collider fades away, and the collision rate decreased to a few Hz. Only the φ meson violated the scaling at the two lowest energies, although measured with very low statistics. Future experiments are required to continue these measurements towards lower beam energies.

3.3.2 Fluctuations of conserved quantities

When approaching the critical point in classical systems such as binary liquids, the density fluctuations increase and the correlation length diverges, finally leading to the phenomenon of critical opalescence. For nuclear matter a similar effect is expected, when varying the beam energy and measuring high moments of the net-baryon number distribution in central heavy-ion collisions. The STAR collaboration reported first evidence of a non-monotonic variation of the 4$^{th}$ order cumulant (kurtosis) times variance of the net-proton number distribution as a function of beam energy, measured event-by-event in central Au+Au collisions [43]. The net-protons number is taken as proxy for the net-baryon number. The fluctuation signal was observed with a 3.1σ significance at $\sqrt{s_{NN}}=7.7$ GeV, the lowest energy measured in the beam energy scan. In contrast, a monotonic variation was found for non-central Au+Au collisions. The fact, that no critical point could be identified at collision energies down to $\sqrt{s_{NN}}=7.7$ GeV, corroborates the prediction of the lattice QCD calculation, that the chiral phase transition temperature is below about 130 MeV, i.e. below collision energies of about $\sqrt{s_{NN}}=6$ GeV. Future high precision measurements at lower beam energies are required, either with STAR in the fixed target mode, or at the upcoming facilities MPD/NICA or CBM/FAIR, in order to discover or to rule out the existence of a first order phase transition with a critical endpoint.

3.3.3 Chemical equilibration of multi-strange hyperons

As discussed above, the yields of all particles measured in heavy-ion collisions at the highest available collision energies can be reproduced by statistical hadronization models, assuming a source with a temperature of about 156 MeV and vanishing baryon chemical potential. This chemical freeze-out temperature corresponds to the pseudo-critical temperature of the crossover transition calculated by lattice QCD. The observation, that also multi-strange (anti-)hyperons are found to be in chemical equilibrium, was interpreted as an indication for a phase transition. It was argued, that because of the small hyperon-nucleon scattering cross section, the hyperons could not reach equilibrium in the short hadronic phase. Instead, multi-strange hyperons are born into equilibrium during the hadronization process, and, therefore, serve as an experimental indication for a phase transition [44]. Chemically equilibrated $\Xi$ and $\Omega$ hyperons were also found in heavy-ion collisions at a beam energy as low as of 30A GeV [45]. At much lower beam energies, however, multi-strange hyperons are not longer equilibrated. While the particle yields measured in Ar + KCl collisions at an energy 1.76A GeV can be explained by the statistical model assuming a chemical freeze-out temperature of $70 \pm 3$ MeV and baryon chemical potential of 748±8 MeV, the $\Xi^-$ hyperon yield overshoots the prediction of the statistical model by a factor of 24±9 [46]. Future experiments are needed, in order to measure precisely the excitation functions of the $\Xi^\pm$ and $\Omega^\pm$ hyperons in heavy collision systems at beam energies between 2 A and 30A GeV, and to search for the equilibration energy for multi-strange (anti-) hyperons, which might indicate the location of a phase transition.

### 3.3.4 Probing deconfinement with charmonium

The melting of J/$\psi$ mesons (c c-bar pairs) in the Quark-Gluon Plasma was one of the first and most promising effects proposed as signature for deconfinement [47]. Later, a charmonium production mechanism was proposed, which was based on the statistical hadronization of charm quarks [48]. The underlying idea was that the charm and anticharm quarks are produced in initial hard collisions, then thermalize together with the other quarks in the QGP, and finally freeze-out into J/$\psi$ and D mesons at the phase boundary. In contrast, the production of J/$\psi$ and D mesons in hadronic collisions like in the Hadron-String-Dynamics (HSD) transport code [49] proceeds via reactions like pp $\rightarrow$ J/$\psi$ pp, pp $\rightarrow$ anti-D $\Lambda_c$ p, and pp $\rightarrow$ D anti-D pp, with threshold energies of $E_{thr}$ = 11.3 GeV, 11.9 GeV, and 14.9 GeV, respectively. Therefore, the yields for J/$\psi$ and D mesons differ for the Statistical Hadronization Model (SHM), where no distinct threshold behavior is taken into account, as compared to the hadronic HSD model, in particular at beam energies close to the production threshold. This effect is illustrated in figure 13, which depicts the (J/$\psi$)/(D+antiD) ratio as function of collision energy, predicted by the HSD code (red dots) for central Au+Au collisions, and by the SHM (blue squares). The ratio increases strongly with decreasing beam energy for the HSD model, because the production threshold for J/$\psi$ meson is lower than for D mesons in hadronic collisions, whereas the ratio predicted by the SHM model increases only weakly. Therefore, the measurement of the J/$\psi$ and D mesons yields in heavy-ion collisions in the beam energy range around the charm production threshold will shed light on the charm production mechanisms, and on the state of matter where charm is created in. Up to now, no charm production has been measured in heavy-ion collisions below top SPS energies.

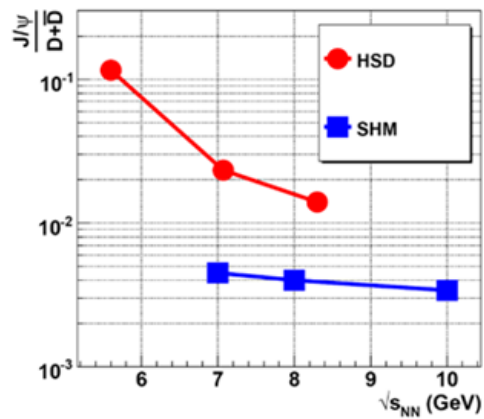

Fig. 13: (J/$\psi$)/(D+antiD) ratio as function of collision energy predicted by the SHM code (blue squares) [48], and by the HSD model (red dots) for central Au+Au collisions [49].

For a hadronic environment, the production of J/ψ and D mesons in heavy-ion collisions may also proceed via the decay of baryonic resonances, which are exited via multiple collisions of baryons and mesons in the fireball. According to UrQMD calculations, a variety of heavy N* resonances are sequentially produced in central Au+Au collisions, which then partly decay via N*→J/ψ+N+N and N*→$\Lambda_c$+anti-D [50]. These decays predominantly contribute to charm production at subthreshold beam energies. Depending on the branching ratio of the decays, yields of $3\times10^{-7}$ J/ψ and $4\times10^{-6}$ $\Lambda_c$ and anti-D can be achieved in a central Au+Au collision at a beam kinetic energy as low as $E_{lab}$= 6 GeV [50]. For a high-rate experiment like CBM at FAIR, it will be possible to accumulate at this beam energy a few thousand of J/ψ mesons per week.

3.3.5 Dileptons: penetrating probes of dense matter

Leptons produced in heavy-ion collisions have the unique feature, that they leave the hot and dense fireball undisturbed by strong interaction. Moreover, lepton pairs as virtual photons provide a direct temperature measurement of the emitting source via the inverse slope of their invariant mass spectrum. This slope, in contrast to the slope of the energy distribution of hadrons, does not include contributions from the radial flow. However, the dilepton invariant mass spectrum up to the φ mass contains contributions from the dilepton decays of vector mesons, which have to be subtracted, in order to extract the temperature of the fireball. Recently the HADES collaboration performed this exercise. They measured the di-electron invariant mass spectrum in Au+Au collisions at a beam kinetic energy of 1.25A GeV, where densities of up to about 2 $\rho_0$ are expected reached. After subtraction of the contributions from known vector mesons, the resulting exponential slope of the invariant mass spectrum could be fitted with a temperature of about 72 MeV [51]. This value represents the average temperature of the fireball, integrated over its evolution. Such dilepton measurements, performed as function of beam energy, would in principle open the possibility to determine the caloric curve of hot and dense QCD matter, and to discover the first order phase transition, if it exist. It is worthwhile to note, that this curve might be attenuated by the small size and the short lifetime of the fireball.

Figure 14 illustrates the fireball temperatures at different stages of its lifetime as function of the collision energy. The purple line, representing the initial fireball temperature $T_{initial}$, and the magenta dashed line, which is the average temperature $T_{slope}$ extracted from the slope of the dilepton invariant mass spectrum in the range between 1–2 GeV/c2, are results of a coarse-graining method applied to an underlying transport model calculation [52]. In this invariant mass range, the spectrum is not polluted by decays of vector mesons. The shape of $T_{slope}$ represents a smooth crossover transition, or the temperature evolution for pure hadronic matter. The black symbols represent the results from HADES [51] and from NA60 [53], which has been extracted from the invariant mass slope above 1 GeV/$c^2$. The blue line represents the freeze-out temperature as shown in figure 12 [40]. The green horizontal line indicates the upper limit of the critical temperature for the chiral transition calculated by lattice QCD in the chiral limit [36]. Given this upper limit, a possible caloric curve should end at this temperature, corresponding to a collision energy below about $\sqrt{s_{NN}}$ = 3 GeV, assuming the fireball temperature defined by the model calculation (magenta line). This would correspond to a net-baryon density of about 3.5 $\rho_0$, and, consequently the region of phase coexistence should be located at even lower densities, which seems, however, not very likely. One explanation for this puzzle might be that QCD matters does not feature a first order phase transition with a critical point, but rather smoothly changes its degrees-of-freedom from hadrons to quarks with increasing temperature and density, as illustrated in figure 11. On the other hand, the beam energy range of interest, i.e. between 2A and 10A GeV, has been poorly studied experimentally up to date, and future facilities like the CBM experiment at FAIR will explore this region in detail, by performing multi-differential measurements of dileptons and hadrons.

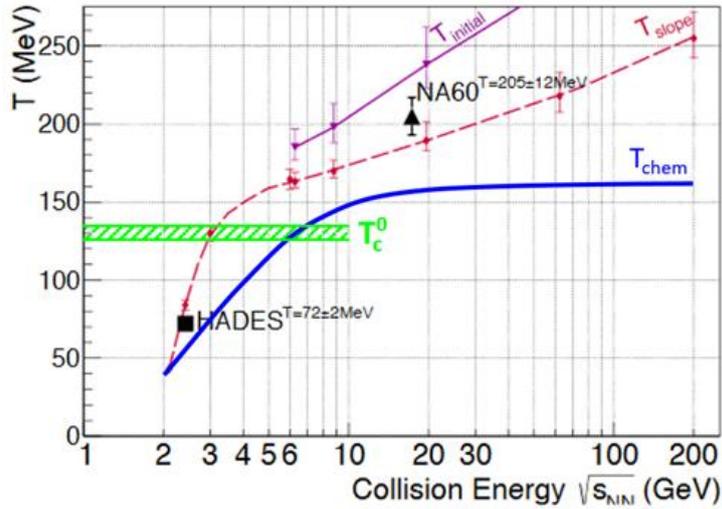

Fig.14: Temperature of the fireball at different stages versus collision energy. Purple line: Initial temperature calculated by applying a coarse-graining method to a transport model [52]. Magenta dashed line: same model calculation, but temperature extracted from the intermediate dilepton invariant mass range (1–2 GeV/c$^2$). Black symbols: Temperature derived from the invariant slope of the dielectron invariant mass spectrum measured by HADES [51], and from the dimuon intermediate invariant spectrum measured by NA60 [53]. Blue solid line: Chemical freeze-out curve as shown in figure 12 [40]. Green area: Maximum critical temperature for the chiral phase transition from lattice QCD [36].

## 4. The Compressed Baryonic Matter experiment at FAIR

The FAIR beam energy range is very well suited to study the high-density EOS and to search for phase transitions in QCD matter, as the net-baryon densities created in central heavy-ion collisions are as high as in neutron star cores. On the other hand, the production cross sections of the most promising diagnostic probes, like the multi-strange (anti-) hyperons and dileptons are extremely low. This is illustrated in figure 15, which depicts the multiplicity of various particles created in one central Au+Au collision at a beam kinetic energy of 4A GeV according to a statistical model calculation. Up to date, no $\Xi$ or $\Omega$ hyperons have been measured at this energy. It should be noted that the typical reconstruction efficiency of multi-strange hyperons is in the order of 1%. Therefore, high multi-differential precision measurements of rare particles in heavy-ion collisions require reaction rates in the order of 1 MHz or above. This is also true for dilepton experiments, because the branching ratios for vector meson decays are in the order of 10$^{-4}$.

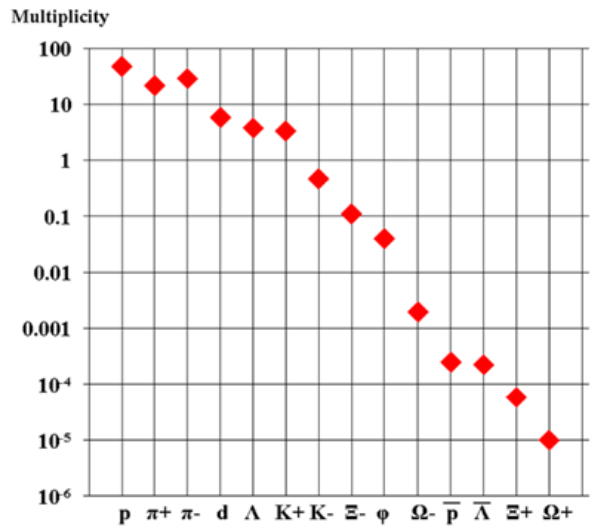

Fig. 15: Multiplicity of produced particles in a central Au+Au collision at a beam kinetic energy of 4A GeV ($\sqrt{s_{NN}}$ = 3.3 GeV) as calculated with a statistical model.

In order to perform high-statistics precision measurements of rare observables, the CBM experiment at FAIR is equipped with fast detectors, which are connected via high-speed readout electronics to a high performance computing centre, where the data are analyzed and selected online. Reaction rates for Au+Au collisions up to 10 MHz are envisaged for selected observables. The CBM detector setup comprises a large aperture superconducting dipole magnet, which hosts a Micro-Vertex Detector (MVD) and a Silicon Tracking System (STS). The MVD consists of 4 detection layers equipped with Monolithic Active Pixel Sensors (MAPS), and will measure with high resolution the decay vertices of short-lived particles. The STS comprises 8 detection layers equipped with double-sided micro-strip sensors. MVD and STS will determine the tracks of charged particles inside the magnetic field over the length of 1 m downstream the target. After the magnet, either a Ring Imaging Cherenkov (RICH) detector for electron identification, or a Muon Chamber (MuCh) System is located. The MuCh comprises 4 hadron absorbers followed by tracking chambers. Downstream the RICH or MuCh a Transition Radiation Detector (TRD) is located, which in case of electron measurements improves the electron identification for higher momenta together with the RICH, or serves as the last tracking station in case of muon measurements. For hadron identification, a 120 $m^2$ Time-of-Flight (TOF) wall is positioned behind the TRD, consisting of Multi-Gap Resistive Plate Chambers (MRPC). The entire detector setup from MVD to TOF covers particle polar emission angles from about 3° to 25°, which is sufficient to cover midrapidity over the FAIR beam energies. The orientation of the reaction plane is determined with the Project Spectator Detector (PSD), a segmented hadronic calorimeter located downstream the TOF wall. A CAD model of the CBM detector setup is shown in figure 16 (right part), together with the already existing High-Acceptance Di-Electron Spectrometer (HADES, left part). HADES has a large acceptance (polar angles from 18° to 85°), and will be used for proton-nucleus collisions and collisions between low-mass nuclei, in order to measure hadrons and di-electrons.

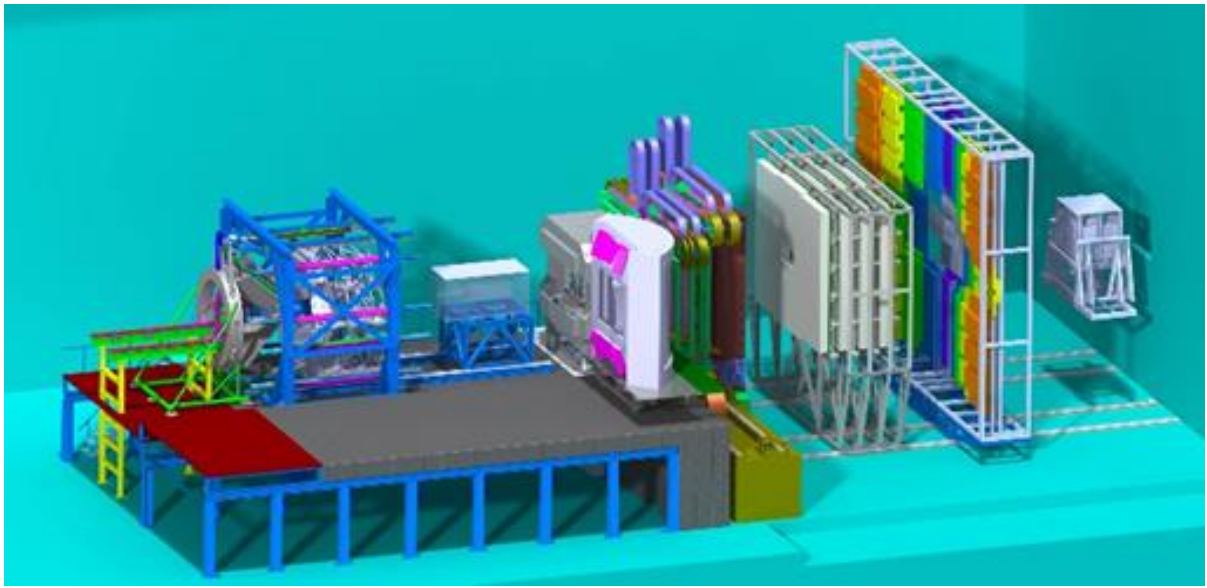

Fig. 16: CAD model of the HADES detector (left) and of the CBM setup (right). The beam enters from left. For HADES operation, the beam will be stopped in the dump in front of CBM. For CBM operation, this dump will be removed, and the beam is focused through HADES onto the CBM target.

As mentioned above, the CBM detector system will be operated at heavy-ion reaction rates of up to 10 MHz, where in each collision hundreds of tracks were recorded. Under such conditions, it is no longer possible to generate a hardware trigger from the track topology of a hyperon decay. To overcome this problem, a trigger-less data read-out and acquisition scheme has been developed. Each individual detector signal will be marked with a time stamp, and sent without any event information to a high-performance computing centre (the "GeenIT cube" of GSI). Here, first the tracks will be reconstructed based on the space and time information of the signals. Then, the particles will be reconstructed, based on the time-of-flight information and track topology. Finally, the events will be reconstructed, based on time information and primary vertex of the tracks. The whole analysis is performed in real-time, which allows to select events containing rare probes by high-speed software algorithms. The resulting performance is illustrated in figure 17, which presents the invariant mass spectra of $K^0_s$, $\Lambda$, anti-$\Lambda$, and $\Xi^-$ reconstructed in min. bias Au+Au collisions at a beam momentum of 10A GeV/c, simulated with the UrQMD generator. In this simulation, the tracks of 300000 events have been distributed according to a Poisson distribution over a time interval of 30 ms, corresponding to a reaction rate of 10 MHz. In table 2, the corresponding total efficiency and the signal-to-background ratio for $\Lambda$, anti-$\Lambda$, and $\Xi^-$ hyperons are listed for an event-based analysis (3-D, upper row), and for 4-D analyses assuming reaction rates of 0.1 MHz, 1 MHz, and 10 MHz.

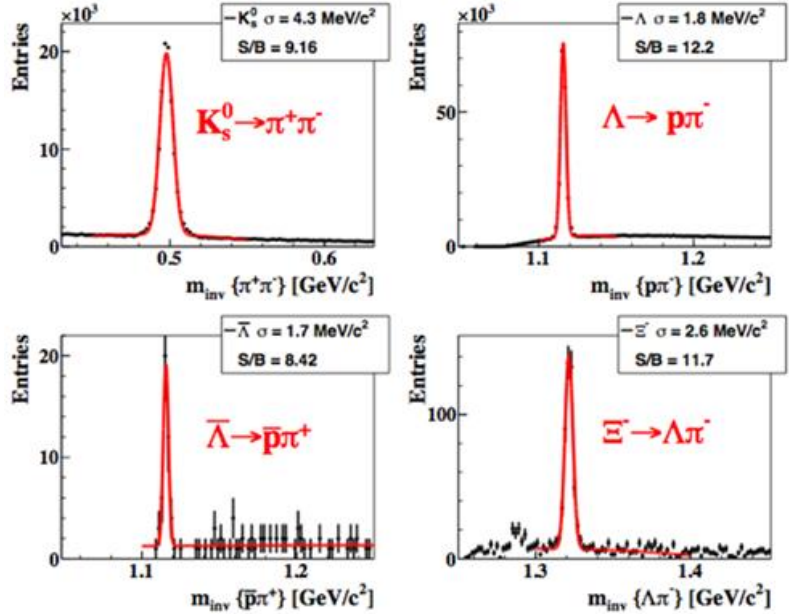

Fig. 17: Invariant mass spectra of $K^0_s$, $\Lambda$, anti-$\Lambda$, and $\Xi^-$ reconstructed in min. bias Au+Au collisions at a beam momentum of 10A GeV/c simulated with the UrQMD generator, assuming a reaction rate of 10 MHz (see text).

Table 2: Total efficiency and Signal-to-Background ratio for $\Lambda$, anti-$\Lambda$, and $\Xi^-$ reconstructed in 300000 min. bias Au+Au collisions at a beam momentum of 10A GeV/c, based on single events (3-D), and for reaction rates of 0.1 MHz, 1 MHz, and 10 MHz (see text).

|         | $\Lambda$ |       | Anti-$\Lambda$ |     | $\Xi^-$ |      |
|---------|-----------|-------|----------------|-----|---------|------|
|         | $\varepsilon/4\pi$ | S/B | $\varepsilon/4\pi$ | S/B | $\varepsilon/4\pi$ | S/B |
| 3-D     | 19.4 | 23.7 | 28 | 12.7 | 10.5 | 21.8 |
| 0.1 MHz | 20.6 | 12.9 | 32 | 10   | 11.7 | 14.2 |
| 1 MHz   | 18.7 | 12.5 | 26 | 10   | 10.6 | 12.3 |
| 10 MHz  | 16.7 | 12.2 | 28 | 8    | 8.2  | 11.7 |

A comprehensive review of the CBM research program is presented in the CBM Physics Book [54]. A brief review of the CBM observables and of simulation results are is given in [55]. A recent report on the status of the development on the CBM detector systems and on various feasibility studies can be found at [55].

# 5. Summary and conclusion

The investigation of fundamental astrophysics questions in laboratory experiments is a fascinating aspect of the research program of FAIR. The superconducting Fragment-Separator and its subsequent experiments will explore the origin of the elements in the universe by the production and investigation of very neutron-rich or neutron- deficient isotopes, which participate in the nucleosynthesis taking place in supernova explosions, in neutron star mergers, or in binaries of a star and a neutron star. The mission of the CBM experiment is to investigate the equation-of-state and the degrees-of-freedom of QCD matter at neutron star core densities. The most promising approach is to study excitation functions of sensitive probes, and to search for characteristic changes of their yields or their phase space distributions. The stiffness of the EOS is reflected in the strength of the collective particle flow, and in the yield of particles produced subthreshold. Both observables are poorly known in the FAIR beam energy range, where the densities are created in heavy-ion collisions, which are relevant for our understanding of neutron stars and neutron star mergers. A change from quark matter to hadronic matter is expected to affect several observables, when decreasing the beam energy from top RHIC energies down to FAIR energies: (i) the quark-number scaling of the elliptic flow should disappear, (ii) the multi-strange hyperons should fall out of chemical equilibration, and (iii) the higher-order fluctuations of the net-baryon multiplicity distributions should exhibit a maximum when reaching the critical endpoint of a first order phase transition. Up to date, none of these effects has been observed down to the lowest beam energy provided by the RHIC collider. One reason might be, that the upper limit of the critical temperature of the chiral phase transition predicted by lattice QCD is only around 130 MeV, which is reached in a fireball created at FAIR beam energies. A very promising observable for the search of a first-order phase transition is the fireball temperature extracted from the dilepton invariant mass spectra at intermediate masses. The excitation function of this temperature might exhibit the reminiscence of a caloric curve, and thus provide a unique and conclusive signature of such a transition. A very smooth temperature distribution might indicate a crossover transition. Again, the relevant energy range to search for these effects will be covered by FAIR.

The CBM detector system is designed to perform measurements of promising observables with high precision and unprecedented statistics. In particular, multi-differential distributions of hadrons including multi-strange hyperons, hypernuclei and dileptons will be investigated, for different collision systems and beam energies. In order to run at reaction rates up to 10 MHz, the setup is equipped with fast and radiation hard detectors and a trigger-less readout electronics. The event reconstruction and selection is performed online by high-speed algorithms, which are tuned to run at multi-core CPUs of a high-performance computing farm. The setup has been optimized by extensive simulation studies, and will be commissioned with beams starting 2025. The realization of FAIR is progressing well, a recent photo of the construction site is shown in figure 18 [57].

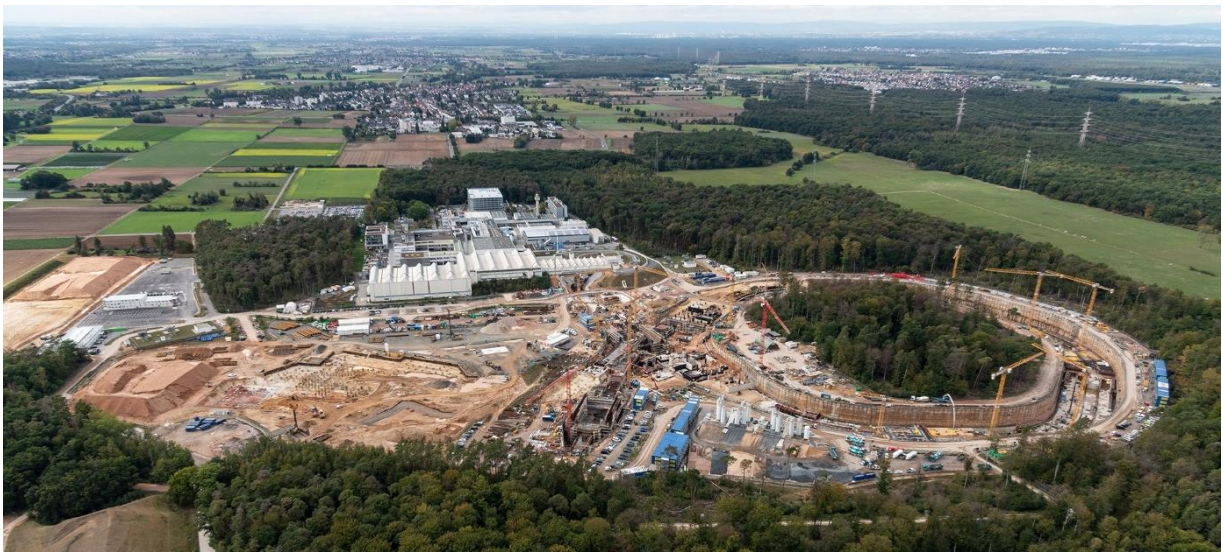

Fig.18: Photo of the FAIR construction site from September 2020 [57].

**Acknowledgement**

The CBM experiment at FAIR is realized by a collaboration of more than 470 scientists from 58 institutions and 12 countries. The project is supported by GSI, the Helmholtz Association, the German Ministry of Education and Research, and by national funds of the CBM member institutions. The project receives funding from the Europeans Union's Horizon 2020 research and innovation programme under grant agreement No. 871072. The author acknowledges support from the Ministry of Science and Higher Education of the Russian Federation, grant N 3.3380.2017/4.6 and by the National Research Nuclear University MEPhI in the framework of the Russian Academic Excellence Project (contract No. 02.a03.21.0005, 27.08.2013).